\documentclass{WileyMSP-template}

\usepackage[cmex10]{amsmath}
\newcommand{\bs}[1]{\boldsymbol{#1}}
\usepackage{ragged2e}
\usepackage{xcolor}
\usepackage{ulem}

\begin{document}

\pagestyle{fancy}

\rhead{\vspace{20pt}
\phantom{Topology optimized plasmonic metasurfaces for optical trapping of nanoparticles} }

\title{Topology optimized plasmonic metasurfaces for optical trapping of nanoparticles}

\maketitle


\author{Emadeldeen Hassan*}



\begin{affiliations}
E.H.\\ 
Department of Applied Physics and Electronics, Ume{\aa}\, 
University, SE-901~87~Ume{\aa}, Sweden\\
emadeldeen.hassan@umu.se
\end{affiliations}



\begin{abstract}
Smart metasurfaces capable of employing the momentum of light for manipulating nanoparticles hold the key to potential applications in science and nanotechnology.
This article proposes a density-based topology optimization framework for optimizing plasmonic metasurfaces for nanoparticles optical trapping.
The Maxwell stress tensor (MST) is employed to compute the optical force exerted on nanoparticles of different sizes and types.
The metasurfaces' topologies are optimized to maximize the gradient (attractive) force on such nanoparticles subject to normally incident monochromatic excitation.
Designs based on free-form optimization are investigated first, then manufacturing constraints are imposed to provide easy-to-manufacture planar designs.
The results show that the topology of the optimized metasurfaces depends on the nanoparticle size and material, with a higher trapping stiffness associated with small nanoparticles.
The optimized metasurfaces could offer selective mass trapping of nanoparticles for applications in biosensing, microfabrication, or assembly of quantum systems.
\end{abstract}


\section{Introduction}
Light-matter interaction at micro- and nanoscale is the basis for many contemporary advances in science and technology.
The exchange of momentum between light and nanoparticles has been exploited to optically manipulate nanoparticles, enabling applications such as optical tweezers\cite{ashkin70Acceleration,Gordon73Radiation,Zhang21plasmonic,Trapping26Trapping}, atoms cooling\,\cite{Phillips98laser}, particles trapping\cite{ashkin78trapping,kim25blinking}, and investigations of nanomotors and laser-induced particle accelerators\,\cite{Doron19Design,Shao18Lightdriven,Serrera25Nanomotor}.
Optical trapping of nanoparticles is typically achieved via two approaches.
The first approach relies on using laser beams to trap or transport nanoparticles in aqueous solutions\,\cite{ashkin70Acceleration}. 
The gradient force pulls nanoparticles toward the high-intensity region at the center of the laser beam, while the scattering force transports them along the direction of propagation.
This approach is particularly useful in biophotonics\,\cite{ji25nanoscale}.
The second approach of optical trapping uses the near-field enhancement on smooth (or nanostructured) surfaces to manipulate nearby nanoparticles\cite{Salary16Tailoring,Righini09Nano,Ploschner10Optical}. 
While the first approach enables the manipulation of nanoparticles in 3D (using multiple laser beams), the second approach is typically used for near-surface 2D nanoparticle trapping.

The near-surface 2D trapping relies on the interaction of light with nanostructures, for example by exciting surface plasmons or enhancing fields at the edges or inside the gap of nanoantennas\,\cite{Novotny97Theory}.
Compared to dielectrics, plasmonic media have the potential to offer high-field enhancement using dimensions much smaller than the diffraction limit\cite{Zhang21plasmonic,Belkin15Plasmonic,kang15Trapping}, which offers high spatial precision when manipulating nanoparticles.
Although they might be criticized for exhibiting dissipative losses that could raise their temperature, the metals density could be controlled to reduce the thermal-induced dynamics\cite{Righini09Nano} or they can be employed to disperse heat as their thermal conductivity is typically high\cite{Xu18Direct}.
Another critical aspect is that the presence of the nanoparticles close to the edges of a nanostructure might detune the resonance state of the configuration, and it is imperative to account for the presence of the nanoparticles during the design of such 2D optical traps\,\cite{juan09self,Ploschner10Optical}.

For particles much smaller than the wavelength, the optical force is proportional to the gradient of the electric-field intensity and the polarizability of the particle\cite{Novotny12Principles}. 
In this case, the field enhancement can be used as an implicit objective for optimizing the near-surface 2D optical force \cite{Nelson24Inverse,Jokisch24Omnidirectional}.
While this objective is justifiable for low index small particles, a more accurate approach is to explicitly optimize the force exerted on the nanoparticle.
This is particularly essential for high-index particles or particles with sizes comparable to the wavelength, where the simple dipole approximation may no longer be valid\cite{Novotny12Principles}. 
Salary et al \cite{Salary16Tailoring} used exact multipole expansion and Maxwell stress tensor (MST) to study optical forces exerted on dielectric and plasmonic spherical nanoparticles located on multilayered substrates subject to Gaussian beam excitation.
Jokisch et al \cite{Jokisch24Omnidirectional} relied on the dipole approximation and proposed a topology optimized omnidirectional optical trap based on a waveguide-integrated dielectric nanocavity, which is effective for trapping single, minute particles. 
However, trapping particles inside a cavity might limit the trap usefulness.
Be\~nat et al \cite{Martinez25Engineering} proposed a two-dimensional inverse-design framework based on the Maxwell stress tensor to simultaneously design particles and a nearby dielectric lens to maximize the scattering force.

In this work, we propose to optimize plasmonic metasurface aiming at trapping near-surface nanoparticles.
The metasurfaces are optimized using a density-based topology optimization framework\,\cite{TopOptGeneral,HaWaBe14,Gedeon25Time,Hassan:22}.
We use the finite-difference frequency-domain (FDFD) method to solve the 3D Maxwell's equations, and the optical force is computed using the MST.
The adjoint-field method is used to compute the objective function gradient, and the distribution of a plasmonic material (gold) is optimized to maximize the gradient force.
To the best of our knowledge, this work is the first to explicitly optimize plasmonic metasurfaces using the MST with the nanoparticles included in the optimization. 
The force is optimized in presence of the nanoparticles, while a thin silica superstrate layer separates the nanoparticle from the plasmonic structure to maintain a smooth surface and avoid detuning the trapping mechanism.
This approach makes the force uniquely dependent on the properties of the metasurface and nanoparticles, making it suitable for selectivity trapping, while accounting for the self-induced back-action\,\cite{juan09self}. 
The plasmonic metasurfaces provide higher gradient-forces and enable a high spatial density of trapping sites, which might be useful in applications such optical tweezer arrays for assembly of quantum systems or large-scale selective trapping of nanoparticles.

\section{Methods} 

\subsection{Optimization setup and problem statement} 
\label{S:ProMod}
\begin{figure} 
\centering
\includegraphics[trim = 3mm 0mm 1mm 0mm, clip,width=0.45\columnwidth,draft=false]{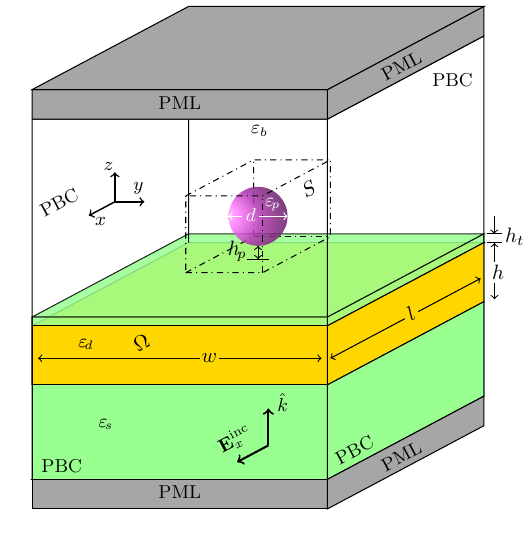}
\caption{Optimization setup. A unit cell of a plasmonic metasurface consisting of a periodic structure in $x$- and $y$-directions. A spherical nanoparticle with permittivity $\epsilon_p$, diameter $d$, centered horizontally inside the unit cell and separated a distance $h_p$ above a layer of silica with thickness $h_t$. 
A design domain $\Omega$ with dimensions of $l\times w\times h$ where a plasmonic material (gold) will be distributed to maximize the force exerted on the particle along the negative $z$-direction subject to an $x$-polarized incident plane wave propagating in the positive $z$-direction. 
The nanoparticle, enclosed by the virtual surface $S$, is immersed in a background medium with permittivity $\epsilon_b$.}
\label{Problem}
\end{figure}

Figure\,\ref{Problem} shows the optimization setup. 
A spherical nanoparticle with a diameter $d$ and a permittivity $\epsilon_p$ is immersed in a medium with permittivity $\epsilon_b$ and horizontally located at the center of a unit cell of a periodic metasurface.
The unit cell has dimensions of $l\times w$ in the $x$- and $y$-directions, respectively. 
A design domain $\Omega$, having the same transverse dimensions as the unit cell and a thickness $h$, is enclosed by silica on both the bottom (substrate) and top (superstrate) sides.
The silica on the top side has a thickness $h_t$, acting as a spacer to prevent the nanoparticle from touching the plasmonic material to prevent permanent binding of the nanostructure to the surface \cite{Belkin15Plasmonic} or detuning the trapping mechanism caused by the charge configurations on both sides.
For the optimization purpose, the position of the nanoparticle is assumed fixed at the middle of the unit cell and located a distance $h_p$ above the silica superstrate.
We aim to distribute a plasmonic (here, we chose gold) material inside the domain $\Omega$ to maximize the attractive force (i.e., the force in the negative $z$-direction) between the plasmonic metasurface and the nanoparticles.
The excitation is an $x$-polarized plane wave of intensity $I_0$, impinging from the silica substrate and propagating in the positive $z$-direction.

\subsection{Objective function: the optical force}
To optimize for optical trapping, we use the electromagnetic (optical) force as a measure.
A charge $q$ exposed to an electric field $\bs{\mathcal{E}}$ and moving with a velocity $\nu$ in a magnetic flux $\bs{\mathcal{B}}$ experiences an electromagnetic force $f = q [\bs{\mathcal{E}} + \nu \times \bs{\mathcal{B}}]$\cite{Griffiths23introduction}. 
Here, we assume time-harmonic fields of the form ${\mathcal{E}} = \Re{\left\{ E e^{j\omega t}\right\}}$.
For a particle immersed in a domain with a permittivity $\varepsilon_b$, by using the force density $f$ and Maxwell's equations,  the time-average optical force exerted on the particle can be expressed as \cite{Barton89Theoretical,Novotny12Principles},
\begin{align}\label{Objective} 
\langle \bs{F} \rangle = & \oint_{S} \langle \overset{\leftrightarrow}{\bs{T}} \rangle \cdot nd\bs{S}  
\end{align}
with $\langle {T}_{\alpha\beta} \rangle  = \frac{1}{2} \Re\left\{ \varepsilon_b  E_\alpha E_\beta^{*} + \textcolor{black}{\mu} H_\alpha H_\beta^{*} -  \frac{1}{2} ( \textcolor{black}{\varepsilon_b} \bs E \cdot \bs{E^{*}} + \textcolor{black}{\mu} \bs H \cdot \bs{H^{*}}) \delta_{\alpha\beta} \right\}$ denotes the time-average Maxwell's stress tensor (MST) and the subscripts $\alpha$ and $\beta$ indicate the coordinates directions $x$, $y$, and $z$; $*$ is the complex conjugation; $\delta_{\alpha\beta}$ is the delta function; the total electric $\bs{E}$ and magnetic  $\bs{H}$ field vectors defined on the enclosing virtual surface $S$ (see Figure\,\ref{Problem}); and $\mu$ is the medium permeability. 
An alternative approach to compute\,\eqref{Objective} is to use the divergence of $\overset{\leftrightarrow}{\bs{T}}$ and integrating over the volume enclosed by $S$, however, this approach requires more computations.

\noindent We obtain the $\bs{E}$ and $\bs{H}$ fields by solving the 3D Maxwell's equations
 \begin{subequations}\label{MaxwellEq}
\begin{align}
j \omega \varepsilon \bs E  + \bs J - \nabla\times \bs{H} = \bs 0 , \label{AmpereEq} \\
j \omega \mu  \bs H+ 	\times\boldsymbol{E} = \bs 0 \label{FardayEq},  
\end{align}
\end{subequations}
with $\bs J$ denoting the electric current source and $\omega$ is the radian frequency.
The $\bs H$ field can be eliminated from\,\eqref{MaxwellEq}, and Maxwell's equations can be re-expressed using the E-field wave equation\cite{jin14FEM},
\begin{align}
\nabla\times \left(\frac{1}{\mu} \nabla\times\boldsymbol{E} \right) -\omega^2  \varepsilon \bs E = -j \omega \bs J. \label{WaveEqI}
\end{align}
We use the finite-difference frequency-domain (FDFD) method to discretize and numerically solve\,\eqref{WaveEqI}\cite{Sharkawy06Plane,Zainud09Electromagnetic}.
The computational domain is discretized using the cubical Yee cell\cite{Taflove05}.
The radiation boundary condition in the $z$-direction is implemented using the uniaxial perfectly matched layer (UPML), and the periodic boundary condition in the $x$- and $y$-directions is imposed using the Floquet theory\cite{jin14FEM}. 
The total electric field is decomposed into incident and scattered fields, and the total-field scattered-field formulation is used to impose the plane wave excitation\,\cite{Zainud09Electromagnetic,rumpf12simple}.
By using the FDFD\,method, the discretized E-field wave equation can be expressed in a matrix form as 
\begin{align}\label{WaveEqDiscI} 
\boldsymbol{A} \bs e  = \bs b 
\end{align}
where $\boldsymbol{A}$ is the system coefficient matrix, the vector $\bs{e} $ holds the unknown electric field components in the discretized Yee grid, and $\bs b $ denotes the excitation.
We assemble the linear system\,\eqref{WaveEqDiscI} and solve it using the QMR iterative method in MATLAB.
We assume the objective function is the $\alpha$-component of the time-average force $\langle F_\alpha(\bs e (\varepsilon))\rangle$.
After solving the linear system \eqref{WaveEqDiscI} and by using the definition of the MST, we numerically estimate the time-average force using 
\begin{align}\label{ObjectiveDiscI} 
\langle \hat{F}_\alpha \rangle & = \frac{\Delta_{s}}{2} \Re\left\{ \textcolor{black}{\varepsilon_b} \bs{e}_{\alpha}^T \bs{e}_{\beta}^{*} + \textcolor{black}{\mu} \bs{h}_{\alpha}^T \bs{h}_{\beta}^{*} - \frac{1}{2}\left( \textcolor{black}{\varepsilon_b} \bs{e}^{T} \bs{M}_{\delta e}  \bs{e}^{*} +\textcolor{black}{\mu} \bs{h}^{T} \bs{M}_{\delta h}\bs{h}^{*} \right) \right\},
\end{align}
where the superscripts $T$ and $*$ denote the transpose and the complex conjugation, respectively. 
The vectors  $\bs{e}_{i}$ and $\bs{h}_{i}$ ($i=\alpha, \beta$) are sub-vectors of the electric and magnetic field vectors $\bs{e}$ and $\bs{h}$, defined on the enclosing surface $S$ that is discretized using a uniform differential surface element $\Delta_{s}$.
We estimate the magnetic field $\bs{h}$ from the computed $\bs{e}$ vector using Faraday's law\,\eqref{FardayEq}.
The sub-vectors $\bs{e}_{i}$ on $S$ can be expressed as $ \bs{e}_{i}= \bs{M}_{ei} \bs{e}$ with $\bs{M}_{ei}$ ($i=\alpha, \beta)$ denote sparse selection matrices.
We can define the sub-vectors $\bs{h}_{i} = \bs{M}_{\!hi} \bs{e}$ via the selection matrices $\bs{M}_{\!hi}$ ($i=\alpha, \beta$).
And similarly, the matrices $\bs{M}_{\delta i}$ ($i=e, h)$ are utilized to estimate the required field components on the surface $S$.
Then, expression\,\eqref{ObjectiveDiscI} can be expressed in terms of the vector $\bs{e}$ as 
\begin{align}\label{ObjectiveDiscIe} 
\langle \hat{F}_\alpha \rangle & = \Re\left\{ \bs{e}^{\dagger} \bs{M}  \bs{e} \right\} \nonumber \\ 
& = \frac{1}{2}\left\{ \bs{e}^{\dagger} \bs{M}  \bs{e} + \bs{e}^{T} \bs{M}^{*}  \bs{e}^{*} \right\}, 
\end{align}
where $\dagger$ denotes the conjugate transpose and $\bs{M}$ is the scaled sum of the selection matrices involved in expression\,\eqref{ObjectiveDiscI}.
We use expression\,\eqref{ObjectiveDiscIe} as our objective function to estimate the optical force. 
We validated our implementation to compute the optical force \eqref{ObjectiveDiscIe} by reproducing the results reported by Ploschner et al. \cite{Ploschner10Optical}.

\subsection{Topology optimization framework} 

We formulate the optimization problem 
\begin{align}\label{OptimizationPro}
 \min_{\varepsilon_{d} \in \{\varepsilon_\text{gold}, \varepsilon_\text{silica} \}}
  \langle \hat{F}_{\alpha} (\varepsilon_{d}) \rangle ,
\end{align}
subject to the governing equations, boundary conditions and a specified illumination. 
Here, $\alpha$ denotes the negative $z$-direction, see Figure\,\ref{Problem}.
Optimization problem\,\eqref{OptimizationPro} aims at minimizing the time-averaged optical force pointing in the negative $z$-direction (i.e. maximizing the attractive force) by distributing the permittivity of a design material (gold) in the design domain $\Omega$, shown in Figure\,\ref{Problem}.
We chose silica as the background material \textcolor{black}{in the design domain} since the design domain is surrounded by silica \textcolor{black}{on both sides}.
However, other background media \textcolor{black}{in the design domain} could also be utilized.

To solve optimization problem\,\eqref{OptimizationPro} using gradient-based optimization methods, the gradient of the objective function is required.
Let $\delta \varepsilon$ be an arbitrary design variation.
By using expression\,\eqref{ObjectiveDiscIe}, we write the first variation of $\langle \hat{F}_\alpha \rangle$ as\,\cite{Wadbro15Topology}
\begin{align}\label{FirstVariation} 
\langle \delta\!\hat{F}_\alpha \rangle & = \frac{1}{2}\left\{ \bs{e}^{\dagger} \bs{M}  \delta\bs{e} + \delta\bs{e}^{\dagger} \bs{M} \bs{e} + \bs{e}^{T} \bs{M}^{*} \delta\bs{e}^{*} + \delta\bs{e}^{T} \bs{M}^{*} \bs{e}^{*}\right\} \nonumber \\
& = \Re\left\{ \bs{e}^{\dagger} \bs{M}_\gamma \delta\bs{e} \right\} 
\end{align}
where the sparse matrix $\bs{M}_\gamma =  \bs{M}+\bs{M}^{\dagger}$. 
Expression\,\eqref{FirstVariation} is not straightforward to estimate as the first variation $\delta\bs{e}$ is not known.
We employ the linear system\,\eqref{WaveEqDiscI} to estimate expression\,\eqref{FirstVariation}.
The first variation of the linear system\,\eqref{WaveEqDiscI} is
\begin{align}\label{WaveEqDiscPerI} 
\delta\!\boldsymbol{A} \,\bs e +   \boldsymbol{A}\,\delta\bs{e} = \bs 0.
\end{align}
By taking the scalar product of\,\eqref{WaveEqDiscPerI} with a vector $\bs{\gamma}$, we obtain 
\begin{align}\label{WaveEqDiscPer} 
\bs{\gamma}^T \boldsymbol{A} \,\delta\bs{e} = - \bs{\gamma}^T \delta\!\boldsymbol{A} \,\bs e.
\end{align}
By choosing $\bs{\gamma}^T \boldsymbol{A} = \bs{e}^{\dagger} \bs{M}_\gamma$, we can use the right side of\,\eqref{WaveEqDiscPer} and express the first variation $\langle \delta\hat{F}_\alpha\rangle$ in expression\,\eqref{FirstVariation} as
\begin{align}\label{Gradient} 
\langle \delta\!\hat{F}_\alpha \rangle & = - \Re\left\{  \bs{\gamma}^T \delta\!\boldsymbol{A} \,\bs{e} \right\} 
\end{align}
where the vector $\bs{\gamma}$ is obtained by solving the so-called adjoint system
\begin{align}\label{AdjointSys} 
\boldsymbol{A} \bs{\gamma} = \bs{M}^{T}_{\gamma} \bs{e}^{*}
\end{align}
and we have relied on the symmetry of $\boldsymbol{A}$. 
In summary, we compute the force by solving the forward system\,\eqref{WaveEqDiscI} and evaluate expression\,\eqref{ObjectiveDiscIe}.  
The objective function gradient is obtained by solving the adjoint system\,\eqref{AdjointSys} and using expression\,\eqref{Gradient}.

We solve problem\,\eqref{OptimizationPro} by using the density-based topology optimization approach\cite{TopOptGeneral,HaWaBe14,Gedeon25Time}. 
In this approach, the design domain is divided into small elements and a density variable $\rho\in\{ 0,1\}$ is associated to each element to indicated presence $(1)$ or absence $(0)$ of the design material.
Therefore, an arbitrary design can be described using the density vector $\bs{\rho} = (\rho_1,\rho_2, \cdots \rho_N)$ with $N$ denoting the number of elements in the design domain.
In the FDFD method, the permittivity is defined on the edges of the Yee cell\,\cite{Taflove05}, and we associate each entry of $\bs{\rho}$ with a corresponding edge in the design domain.
To solve optimization problem\,\eqref{OptimizationPro} by using gradient-based optimization, each design variable $\rho_i$ is allowed to vary continuously between $0$ and $1$ during the optimization.
We use the nonlinear mapping scheme, proposed by Christensen et al \cite{christiansen19nonlinear}, to relate the permittivity and the density variable.
In this mapping, the complex design permittivity at edge $i$ is expressed as 
\begin{align}
\varepsilon_{d}(\rho_i) &= \varepsilon'-  j \varepsilon'' ,
\end{align}
where 
\begin{align}
\varepsilon' &= n^2 -k^2,  && \varepsilon'' = 2nk,   \\
n & = n_{bk} + \rho_i (n_{d}-n_{bk}), && k  = k_{bk} + \rho_i (k_{d}-k_{bk}), 
\end{align}
with $n$ and $k$ denote the real and imaginary parts of the complex refractive index, while the subscripts $d$ and $bk$ refer to the design (gold) and background (silica) material, respectively.
The nonlinear mapping introduces artificial damping for intermediate designs, which is essential to avoid numerical convergence issues when optimizing plasmonic media\,\cite{christiansen19nonlinear,Hassan:22,Gedeon25Time}.

To avoid fast convergence to low-performing designs or designs with checkerboard patten, the design variables are typically filtered\,\cite{Si07,SvSv13,HaWa16}. 
Here, we employ a two-phase nonlinear filter\,\cite{Hassan2018}, and the filter parameters are updated following a continuation approach to ensure convergence to binary designs (i.e., designs with $\rho_i\in \{ 0,1\}$ for all $i$).
The derivatives of the objective function with respect to the filtered design is obtained using the chain rule\,\cite{Hassan:22}.
We use the globally convergent method of moving asymptotes (GCMMA)\,\cite{SvanbergGlobally} to update the design variables and solve problem\,\eqref{OptimizationPro}.

\begin{figure} 
\centering
\includegraphics[trim = 0mm 5mm 0mm 5mm, clip,width=0.65\columnwidth,draft=false]{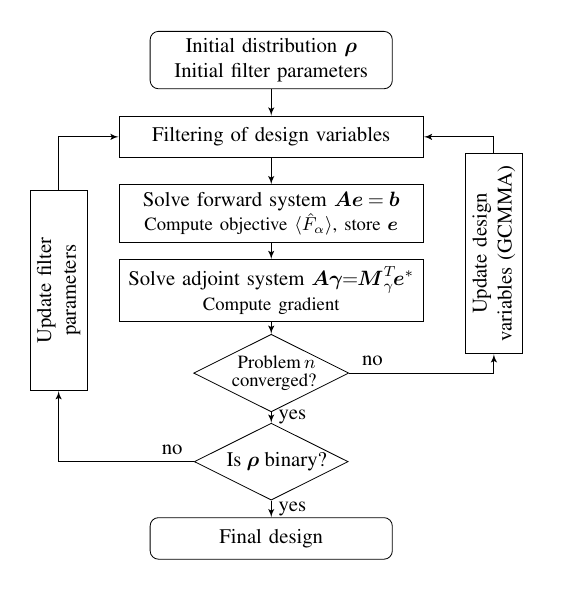}
\caption{A flowchart of the optimization framework.}
\label{FigFlowChart}
\end{figure}

The flowchart of the optimization framework is presented in Figure\,\ref{FigFlowChart}.
For a given set of filter parameters, the sup optimization problem $n$ is solved iteratively until a convergence criterion is satisfied. 
In the inner loop of the design algorithm, we employ a stopping criterion based on a 50\%\,reduction in the norm of the first-order optimality condition compared to an initial value, recorded after eight iterations when solving problem $n$, or upon reaching a maximum of $25$-iterations.
In the outer loop, we monitor the designs non-discreteness\,\cite{Si07}, and the algorithm terminates when it falls below 0.5\% or when the total number of iterations exceeds $600$.

\section{Results}

\begin{figure}[]
\centering
\includegraphics[trim = 0mm 0mm 2mm 2mm, clip,width=1.0\columnwidth,draft=false]{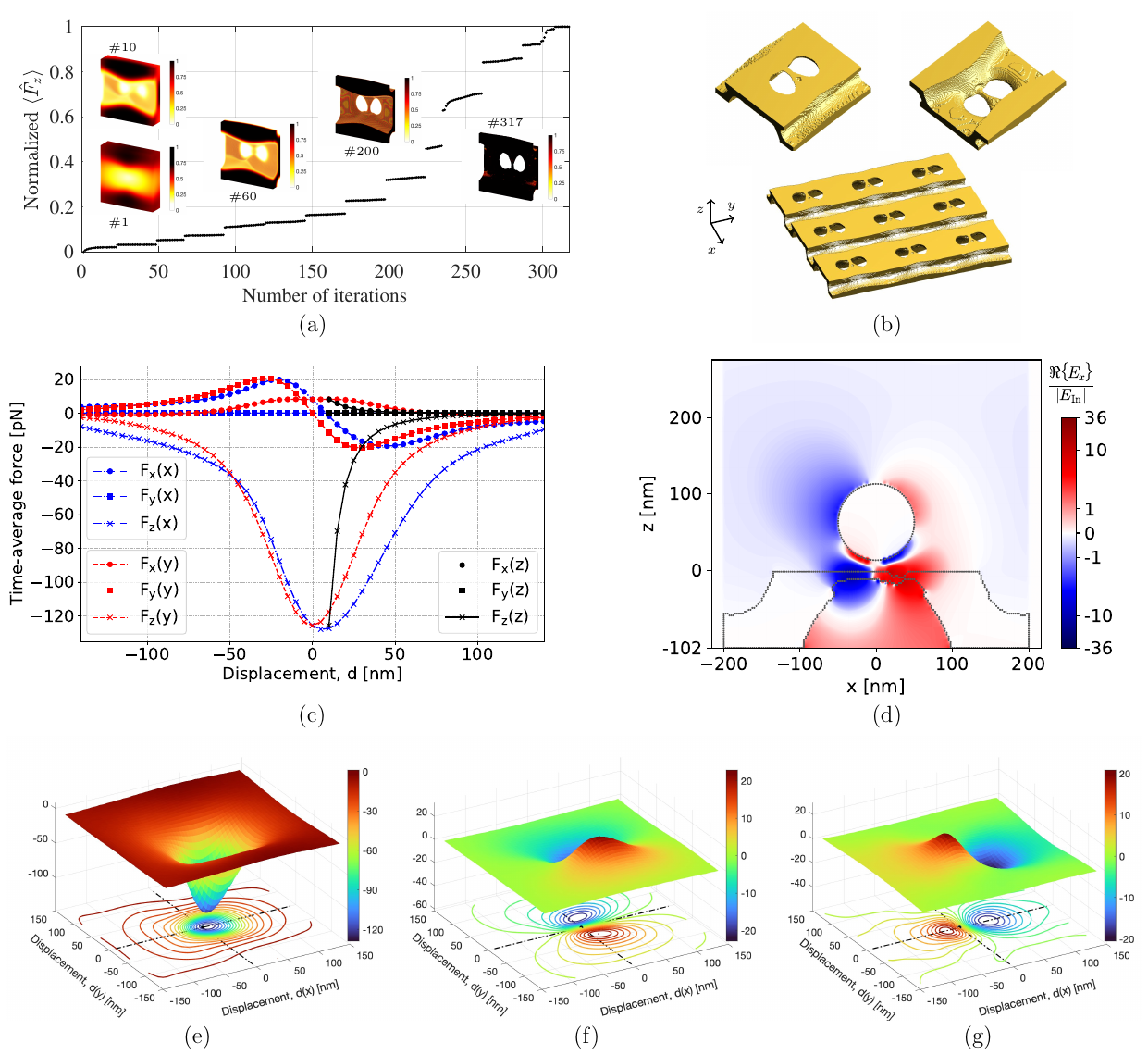}

\caption{A plasmonic metasurface optimized with a 100\,nm-diameter gold nanoparticle placed 10\,nm above the superstrate. (a) Evolution of the normalized objective function and snapshots showing the development of the design. (b) Top and bottom views of the final design and a demonstrative $3\times3$ unit cells of the metasurface.  (c) The force components as a function of the nanoparticle displacement along the major Cartesian axes. Point $0$ denotes the reference point, located at the middle of the unit cell, 10\,nm above the 5\,nm-thick superstrate, see Figure\,\ref{Problem}. 
(d) An $xz$-cross section view through the middle of the unit cell, showing the real-part of the \textcolor{black}{normalized} $x$-component of the electric-field. The attractive force is caused by two dipole modes; one on the metasurface and one on the nanosphere, exhibiting opposite polarities (see Movie\,1). (e), (f), and (g), respectively, surface plots of the time-average forces $F_z$, $F_y$, and $F_x$ as functions of the transverse displacement when the nanoparticle is located 10\,nm above the $5$\,nm superstrate. 
}
\label{FigObjective}
\end{figure}

We use the setup shown in Figure\,\ref{Problem} and solve optimization problem\,\eqref{OptimizationPro} to maximize the attractive force between spherical nanoparticles and gold metasurfaces. 
\textcolor{black}{The background medium enclosing the nanoparticle is assumed to be water with relative permittivity $\epsilon_b= 1.76$, and 2.14 is used as the silica relative permittivity.} 
\textcolor{black}{The complex permittivity of gold at the operational wavelength is $\varepsilon_{gold} = -48.90-j3.66$, obtained by interpolating experimental data by Johnson and Christy\,\cite{Au_Johnson}.}
In our simulation, we use a uniform spatial discretization step of $2.5$\,nm, which provides sufficient accuracy at the excitation wavelength $\lambda=1064\,$nm considered in this work. 
The intensity of the incident wave in the computational domain is fixed to $I_0 = 17$\,mW/\textmu m${}^2$\cite{Ploschner10Optical}.
We use a square unit cell with transverse sizes of $w=l=400$\,nm, and the design domain thickness is $h = 100$\,nm, see Figure\,\ref{Problem}. 
The computational domain is terminated by a $12$-cell UPML layers in the $z$-directions.

We start the optimization using an initial density vector with $\rho_i=0.5$ for all $i$, excluding a small aperture of size $8\times8\times40$\,cells at the design domain center, where  $\rho_i=0$ is used.
This aperture ensures a reasonable initial sensitivity since the force is estimated on the side opposite to the direction of the impinging wave. 
Figure\,\ref{FigObjective}(a) shows the history of the normalized objective function and snapshots illustrating the design evolution as a function of the iteration numbers, for the case of a 100\,nm-diameter gold nanosphere. 
The optimization problem includes 3\,064\,320 design variables, and the optimization algorithm converged after 317\,iterations to the final design demonstrated in Figure\,\ref{FigObjective}(b). 
The bottom side, facing the incident wave, includes a tapered structure that ends with  two lung-like apertures separated by a small golden bridge. 
The design is almost symmetric along the $y$-direction \textcolor{black}{whereas} it exhibits a slight \textcolor{black}{asymmetry} along the $x$-direction, caused by the presence of the nanoparticle near the structure during the optimization. 
This issue will be investigated further below.

Figure\,\ref{FigObjective}(c) shows the components of the force, exerted on the nanoparticle, as a function of the particle's displacement relative to a reference point.
The reference point (i.e., point $0$) is defined at the center of the unit cell, separated a distance $h_{p}=10$\,nm above the $5$\,nm-thick silica superstrate, see Figure\,\ref{Problem}.
The positive magnitude of the force indicates that the nanoparticles experience a repulsive force, and the negative magnitude denotes an attractive force.
The variation of the force components along the $y$-direction is almost symmetric (similar to the structure) while the variation along the $x$-direction has its symmetry point shifted around $7$\,nm from the reference point.
The magnitude of the $z$-component of the force decreases exponentially as the nanoparticle moves away from the reference point along the $z$-direction.
We have used only the $z$-component of the force as our objective function, and the unit cell exhibits a trapping mechanism.
That is, the transverse forces around the reference point act in opposite directions, while the normal force has its maximum magnitude acting downward.
Figure\,\ref{FigObjective}(e-g) show surface plots of the force components at the plane $z=10$\,nm, (i.e., $10$\,nm above the $5$\,nm-thick silica superstrate).
The figures confirm that the forces are symmetric along the $y$-axis, with the symmetry point along the $x$-axis located $7$\,nm from the reference point.

To gain insight into the physical mechanism underlying the attractive force between the nanoparticle and the metasurface, we plot the real part of the $x$-component of the electric field in the $y=0$ plane in Figure\,\ref{FigObjective}(d) (see Movie\,1). 
Based on the field plot, we observe that the plane wave excites a dipole mode in the metasurface nanostructure, which in turn induces a dipolar response of opposite polarity in the nanoparticle, resulting in an attractive force.

\subsection{Impact of nanoparticle material and size}

\begin{figure} 
\centering
\includegraphics[trim = 2mm 0mm 4mm 0mm, clip,width=1.0\columnwidth,draft=false]{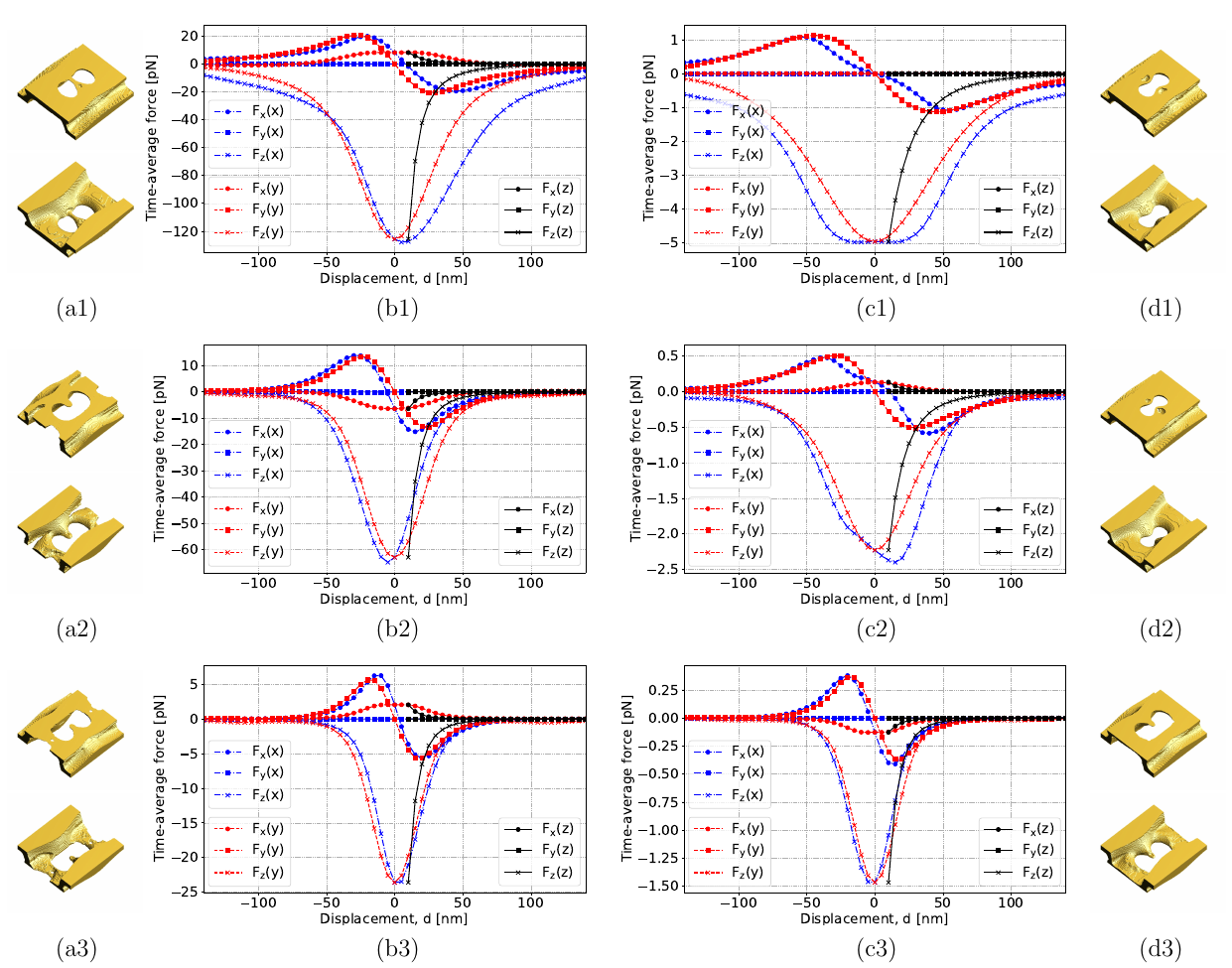}
\caption{Metasurface designs optimized for nanoparticle of different sizes and material types. The first, second, and third rows correspond to nanoparticles with diameters $100$\,nm, $50$\,nm, and $25$\,nm, respectively. The first and second columns are the design results when gold nanoparticle are used, while the third and fourth columns are for dielectric ($\varepsilon_r = 2.25$) nanoparticles. All designs are optimized under identical optimization setup.}
\label{FigMaterialSize}
\end{figure}

We study the impact of the nanoparticle material and size on the topology of the optimized metasurfaces. 
Figure\,\ref{FigMaterialSize} shows plasmonic metasurfaces optimized for plasmonic (gold) and dielectric ($\epsilon_r=2.25$) spherical nanoparticles with diameters of $100$, $50$, and $25$\,nm. 
The designs shown in the left (right) column are optimized considering the presence of gold (dielectric) nanoparticles.
The second column to the right (left) shows the computed force components as functions of the displacement from the reference point.
Qualitatively, we observe that as the diameter of the nanoparticle decreases, the force magnitude decreases, the slope of the curves (also known as the trapping stiffness $k=\frac{\partial F_{\alpha}}{\partial d}$) increases, the structures changes from the popular nanohole/bow-tie shapes to structures with sharp tips\,\cite{Novotny97Theory}.
In all cases, the variation of the force in the $y$-direction is symmetric, whereas the force variation along the $x$-direction (i.e., the polarization direction) exhibits asymmetries in most cases.
The magnitude of the force exerted on the plasmonic nanoparticles is nearly an order of magnitude greater than dielectric nanoparticles of the same size, due to the higher polarizability of gold compared to the dielectric material. 
For both gold and dielectric nanoparticles, the force magnitudes decrease proportional to the decrease in the nanoparticle diameter.

\begin{figure} 
\centering
\includegraphics[trim = 0mm 0mm 0mm 0mm, clip,width=1.0\columnwidth,draft=false]{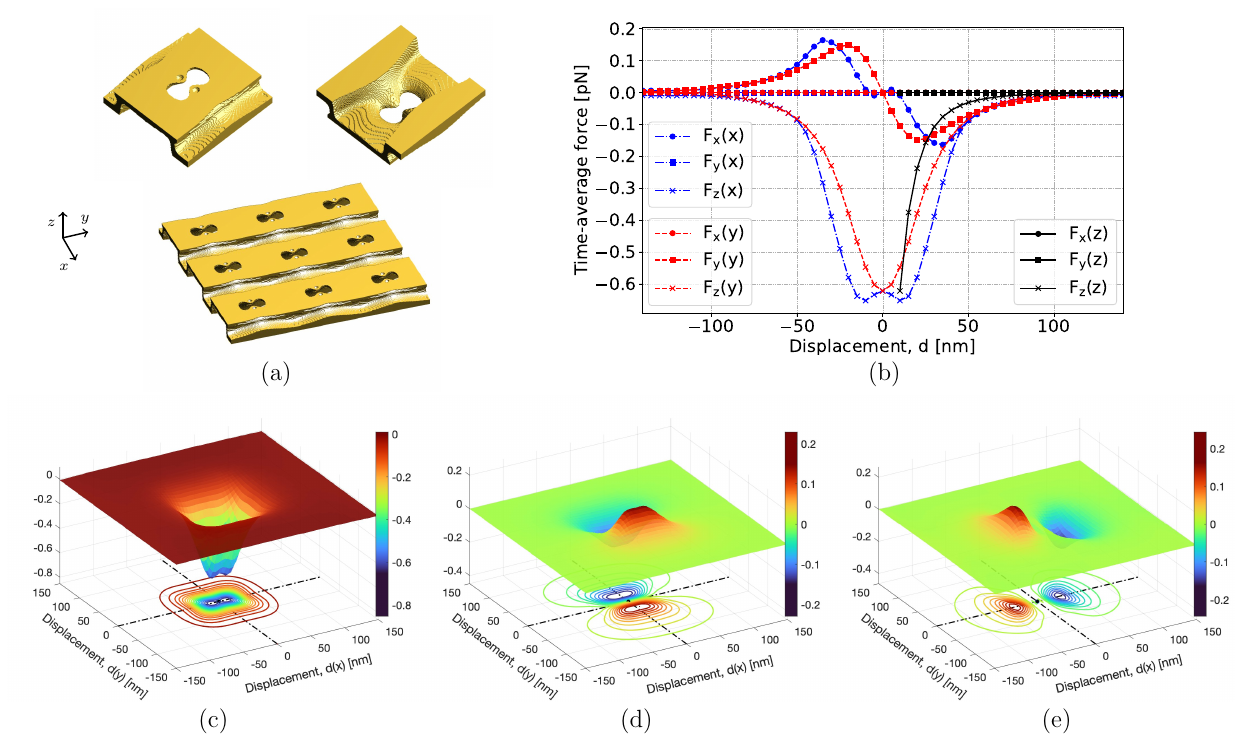}
\caption{Symmetrization of the optical trap by imposing structural mirror symmetry along the polarization direction. (a) Design optimized in the presence of a 25\,nm-diameter dielectric nanoparticle, similar to the case shown in Figure\,\ref{FigMaterialSize}(c3,d3). (b) Force components along the major Cartesian axes, with the origin located at the unit cell center, $10$\,nm above the 5\,nm-thick superstrate. (c), (d), and (f) surface plots of the time-averaged $F_z$, $F_y$, and $F_x$, respectively. By imposing symmetry, the magnitude of the force is reduced by a 60\% compared to the asymmetric design presented in Figure\,\,\ref{FigMaterialSize}(c3)-(d3).}
\label{SymD2_5}
\end{figure}

We investigate the symmetry of the optical force within the unit cell, aiming to make the transverse forces $F_x$ vanish at the reference point.
To symmetrize the trap, this requirement can be addressed either explicitly, by formulating a multiobjective optimization problem that minimizes the transverse forces, or implicitly, by imposing structural symmetry on the design.
Here, we chose the later approach and constrained the design to be symmetric with respect to the $x$-axis. 
Figure\,\ref{SymD2_5} shows a metasurface optimized for a 25\,nm spherical dielectric nanoparticle, with mirror symmetry imposed along the $x$-axis. 
The structure and the computed forces exhibit symmetry along the $x$-axis and $y$-axis.
However, we observe around 60\%\,reduction in the force peak magnitudes compared to the asymmetric design presented in Figure\,\ref{FigMaterialSize}(c3,d3).
That is, the asymmetric design produces higher force magnitudes and greater trapping stiffness compared to the symmetric design.
This tendency for improved performance with asymmetric designs is consistently observed in our numerical investigations, particularly for plasmonic particles, as discussed further below.

\begin{figure} 
\centering
\includegraphics[trim = 2mm 0mm 8mm 0mm, clip,width=1.0\columnwidth,draft=false]{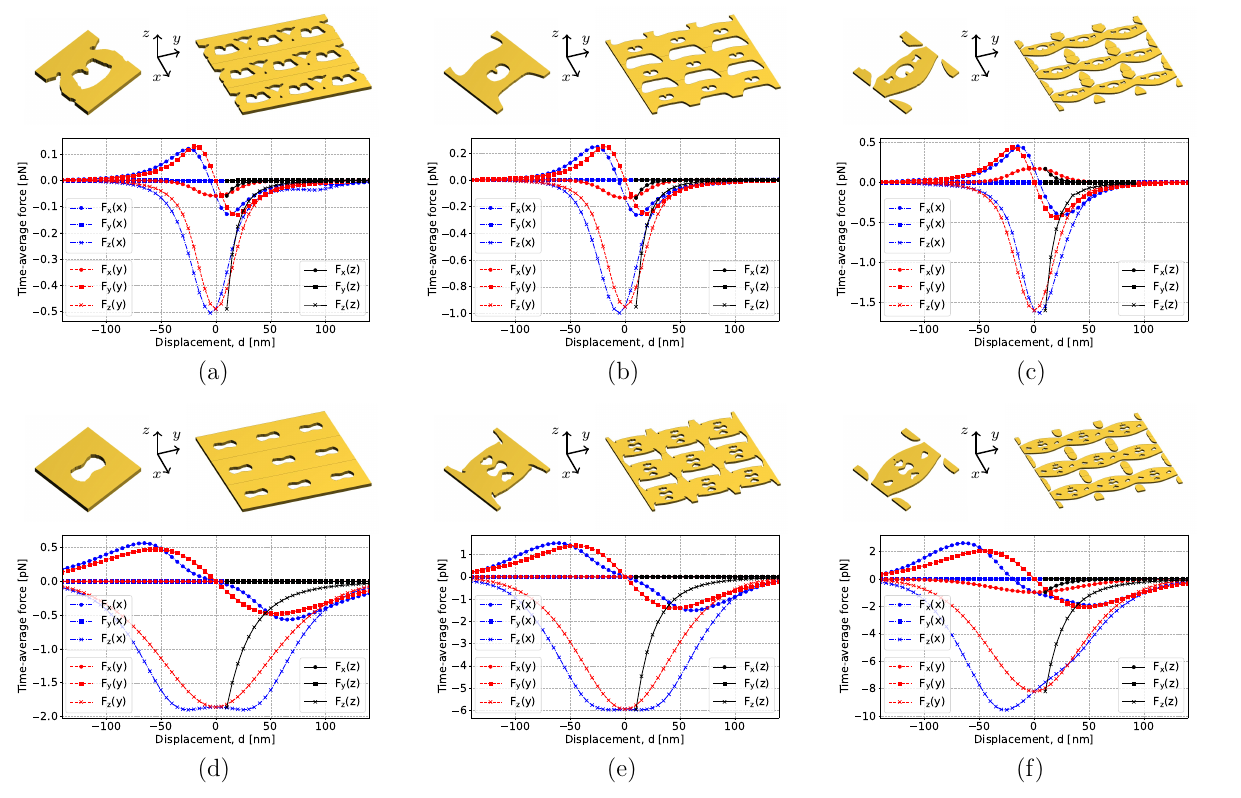}
\caption{Imposing planar invariance on the optimized designs while varying the design domain size. The first row shows designs optimized for trapping a $25$\,nm-diameter dielectric nanosphere, and the second row for a $100$\,nm-diameter nanosphere. The first, second, and third columns correspond to designs domain sizes of $400\times400\times 30$\,nm${}^3$, $600\times600\times 30$\,nm${}^3$, and $800\times800\times 30$\,nm${}^3$, respectively.}
\label{DomainSizeThinD2_5}
\end{figure}

\subsection{Planar invariant designs and impact of unit cell size}

\begin{figure} 
\centering
\includegraphics[trim = 2mm 0mm 8mm 0mm, clip,width=1.0\columnwidth,draft=false]{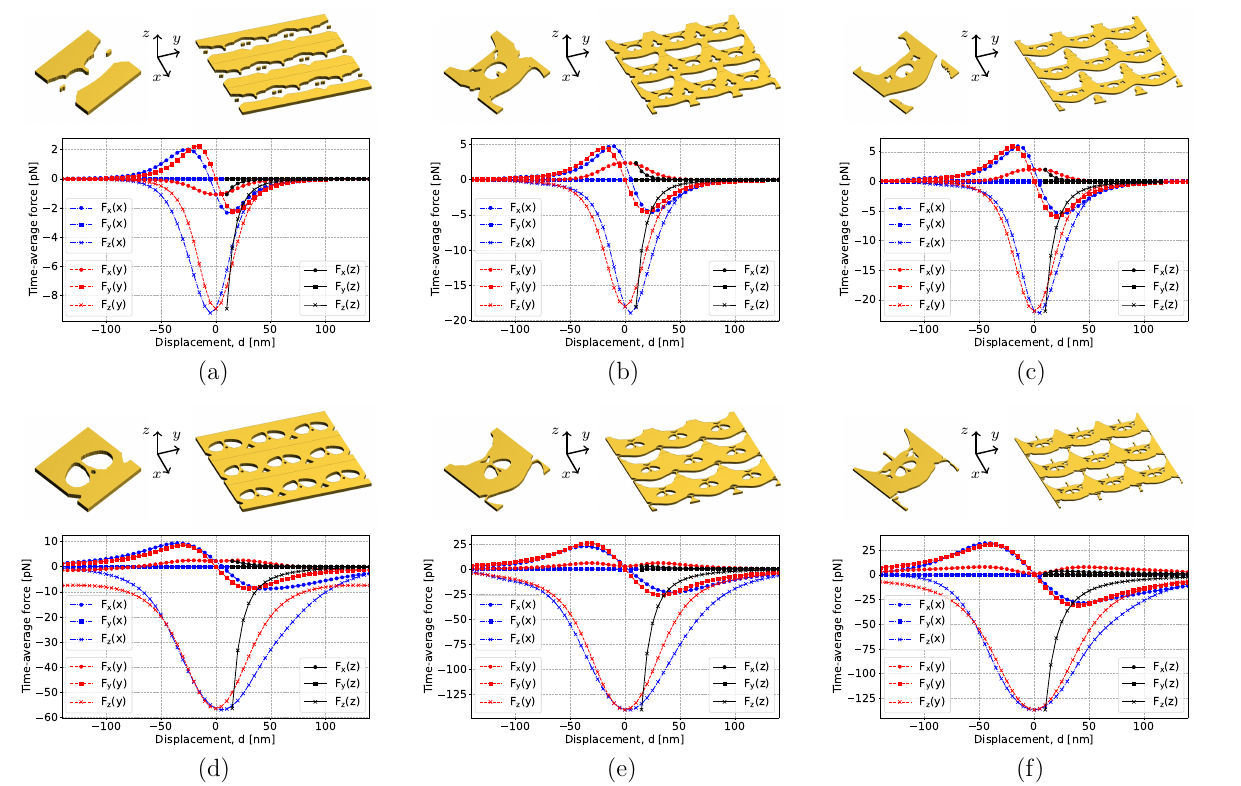}
\caption{Imposing planar invariance on the optimized designs while varying the design domain size. The first row shows designs optimized for trapping a $25$\,nm-diameter gold nanosphere, and the second row for a $100$\,nm-diameter gold nanosphere. The first, second, and third columns correspond to designs domain sizes of $400\times400\times30$\,nm${}^3$, $600\times600\times30$\,nm${}^3$, and $800\times800\times30$\,nm${}^3$, respectively.}
\label{DomainSizeThinA2_5}
\end{figure}

The optimized metasurfaces presented in the previous sections are challenging to fabricate, as they require three-dimensional structuring.
Nevertheless, they are valuable for providing insight into the structural features that emerge during the optimization process. 
To obtain manufacturable designs, we impose planar symmetry on the optimized metasurfaces, which we achieve by applying an averaging filter of the design variables along the $z$-axis. 
For this study, the thickness of the design domain is reduced to $h=30$\,nm.
This choice is motivated by the free-form designs presented in Figure\,\ref{FigMaterialSize} and Figure\,\ref{SymD2_5}, where all structures exhibit slight tapering in the direction of plane-wave incidence, whereas the gold layers tend to be structurally thinner on the side facing the nanoparticle.

Figure\,\ref{DomainSizeThinD2_5} shows planar designs optimized for trapping dielectric nanoparticles with diameters of $25$\,nm (first row) and $100$\,nm (second row).
In the same figure, we also present the effect of varying the unit cell size on the obtained designs. 
The first, second, and third columns correspond to design domain sizes of $400\times400\times 30$\,nm${}^3$, $600\times600\times 30$\,nm${}^3$, and $800\times800\times 30$\,nm${}^3$, respectively.
Similar to the free-from designs, metasurfaces optimized to trap a 25\,nm-diameter nanoparticle exhibit an aperture with a sharp central tip aligned with the incident field polarization and evolving beneath the nanoparticle position. 
In contrast, metasurfaces optimized for trapping a $100$\,nm-diameter nanoparticle feature the characteristic of bow-tie aperture, with the magnitude of the trapping force increase proportionally with the nanoparticle diameter.
As the unit cell size increases, the optimized designs develop additional structural features, contributing an increase in the trapping force magnitude, which is generally proportional to the unit cell surface area.

Figure\,\ref{DomainSizeThinA2_5} shows a similar study for optimizing planar metasurfaces for trapping gold nanoparticles. 
We observe the evolution of design features similar to those seen for dielectric nanoparticles, with two notable differences: (1) the force magnitude is approximately $20$\,times higher than that for dielectric nanoparticles, and (2) designs optimized using the $800\times800\times30$\,nm${}^3$  domain size exhibit force amplitudes comparable to those optimized using the $600\times600\times30$\,nm${}^3$ domain.
In addition, similar to the free-from designs, the metasurfaces optimized for $100$\,nm-diameter nanospheres evolve into two lung-like apertures separated by a golden bridge.

\section{Conclusion}
This work presents a density-based topology optimization framework for designing plasmonic metasurfaces for the optical trapping of nanoparticles. 
By accounting for the presence of the nanoparticles during the optimization, the designs are tailored to the material and size of the nanoparticles.
Our results show that, for small nanoparticles, asymmetric structures with sharp tips aligned along the polarization direction exhibit larger trapping forces compared to symmetric designs, such as the commonly used bow-tie shape. 
For the trapping of small nanoparticles, the optimized structures possess similar topologies regardless of their material type, as they predominantly support electric dipole modes. 
However, for larger nanoparticles the topologies of the optimized metasurfaces depend on the nanoparticle materials. 
As the size of the unit cell increases, more geometrical features evolve in the optimized structures and the magnitude of the trapping force increases. 
The analysis of the field distribution indicates that the attractive force arises from the excitation of two dipolar modes with opposite polarities in the metasurface and the nanoparticle. 
The proposed metasurfaces can be employed for selective mass trapping of nanoparticles in applications such as biosensing, microfabrication, and the assembly of large-scale quantum systems.

\section*{Disclosures} The author declares no conflicts of interest



\medskip
\textbf{Acknowledgements} \par 
The computations were performed on resources provided by the Swedish National Infrastructure for Computing (SNIC) at HPC2N center. The author acknowledges the Kempe Foundation for funding project number:  JCSMK23-0115.

\medskip

%

\end{document}